\documentclass[twocolumn,showpacs,preprintnumbers,amsmath,amssymb]{revtex4}

\usepackage{epsfig}
\usepackage{dcolumn}
\usepackage{bm}

\begin{document}

\title{Braiding of anyonic quasiparticles in the charge transfer statistics of symmetric fractional edge-state Mach-Zehnder interferometer}

\author{Vadim V. Ponomarenko}
\affiliation{Center of Physics, University of Minho, Campus
Gualtar, 4710-057 Braga, Portugal}
\author{Dmitri V. Averin}
\affiliation{Department of Physics and Astronomy,
University of Stony Brook, SUNY, Stony Brook, NY 11794}

\date{\today}


\begin{abstract}
We have studied the zero-temperature statistics of the charge transfer between the two edges of Quantum Hall liquids of, in general, different filling factors, $\nu_{0,1}=1/(2 m_{0,1}+1)$, with $m_0 \geq m_1\geq 0$, forming Mach-Zehnder interferometer. General expression for the cumulant generating function in the large-time limit is obtained for symmetric interferometer with equal propagation times along the two edges between the contacts and constant bias voltage. The low-voltage limit of the generating function  can be interpreted in terms of the regular Poisson process of electron tunneling, while its leading large-voltage asymptotics is proven to coincide with the solution of kinetic equation describing quasiparticle transitions between the $m$ states of the interferometer with different effective flux through it, where $m\equiv 1+m_{0}+m_{1}$. For $m>1$, this dynamics reflects both the fractional charge $e/m$ and the fractional statistical angle $\pi /m$ of the tunneling quasiparticles.  Explicit expressions for the second (shot noise) and third cumulants are obtained, and their voltage dependence is analyzed.
\end{abstract}


\maketitle

\section{Introduction}

Electronic Mach-Zehnder interferometer (MZI) \cite{mz1,mz01,mz02} can be realized with the edge states of the Quantum Hall liquids (QHLs). Together with the quantum antidots \cite{ant,any}, MZIs in the regime of the Fractional Quantum Hall effect (FQHE) are expected \cite{mz3,mz4,usprl} to be useful for observation of the fractional statistics of FQHE quasiparticles. In contrast to fractional quasiparticle charge, which has been confirmed in several experiments \cite{ant,n1,n2}, there is no commonly accepted observation of anyonic statistics of the quasiparticles, which remains a challenging experimental problem. Currently, this problem attracts interest in the context of solid-state quantum computation, since individual manipulation of anyonic quasiparticles involving their braiding provides an interesting possible basis for implementation of the quantum information processing \cite{comp1,comp2,comp3}. However, in typical interferometer-based experimental set-up the quasiparticles emerge as a  continuum of gapless edge excitations that should be described by a 1D field theory \cite{wen}. Individual quasiparticles can be realized in this theory only asymptotically in a special limit. In the fractional edge-states
MZI, such a limit occurs at \emph{large} voltages, when the system is
characterized by the Hamiltonian dual to the Hamiltonian of the initial electron tunneling model of the MZI. The latter is perturbative in electron tunneling at low voltages, and is much better defined, since weak electron tunneling is probably the most basic process in solid-state physics.

In the dual description derived from the electronic model by the instanton technique \cite{usprl,usprb}, MZI acquires $m=m_0+m_1+1$ different quantum states which differ by the the effective flux $\Phi$ through it. This  flux
contains, in addition to the external flux $\Phi_{ex}$, a statistical
contribution earlier also found \cite{usprb05} for the antidots. The tunneling of each quasiparticle changes this effective flux by $\pm \Phi_0$ and therefore switches the MZI from one flux state into another. Since
quasiparticle also carries the charge $e/m$, the change of flux by $\pm \Phi_0$ results in the change of the interference phase for the quasiparticles by $2\pi/m$ and the corresponding change of the rate of coherent tunneling through the interferometer. Summation over the $m$ flux  states in calculation of the physical quantities restores $\Phi_0$ periodicity of their $\Phi_{ex}$ dependence. At low voltages, this periodicity is guaranteed by $\Phi_0$ periodicity of the  electron tunneling amplitude.

In this work, we follow the approach to the MZI that allows us to obtain the  uniform description of its transport properties in both regimes of electron and quasiparticle tunneling. We consider the standard MZI geometry (Fig.~1) with two tunneling contacts between the two effectively parallel edges of QHLs, but allow for, in general, different filling factors, $\nu_{0,1}=1/(2 m_{0,1}+1)$, with $m_0 \geq m_1\geq 0$, of these edges. In the symmetric case of equal propagation times along the two edges between the contacts, the corresponding 1D field theory permits an exact Bethe ansatz solution \cite{usprl}. Making use of this solution, we calculate the zero-temperature full counting statistics \cite{fcs} of the charge transferred between the two edges forming MZI. The transferred charge distribution is shown to reflect the anyonic braiding statistics of the tunneling quasiparticles, and decomposition of electrons into quasiparticles with increase of voltage.

\begin{figure}[htb]
\setlength{\unitlength}{1.0in}
\begin{picture}(3.,1.4)
\put(0.6,-0.05){\epsfxsize=1.9in \epsfbox{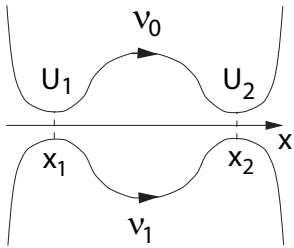}}
\end{picture}
\caption{Mach-Zehnder interferometer considered in this work: two
point contacts with tunneling amplitudes $U_j$ formed at points
$x_j$, $j=1,2$ between two co-propagating edges of QHLs with
different filling factors $\nu_0$ and $\nu_1$. The edges are
assumed to support one bosonic mode each, with arrows indicating
direction of propagation of these modes.}
\end{figure}

The paper is organized as follows. In Section 2 we
introduce the model of symmetric MZI and its Bethe
ansatz solution. In Section 3, we use this solution to
derive the general expression for the generating
function $P(\xi)$ of the distribution of the transferred
charge. The method employed in this calculation
generalizes to the MZI transport through two point
contacts the method developed earlier by Saleur and
Weiss \cite{sw} for a single point contact. The charge
transfer statistics found for one contact \cite{sw}
demonstrated fractionalization of electron charges with
increasing bias voltage across the contact. In the case
of MZI, one can  expect that in addition to charge
fractionalization, braiding properties of anyonic
quasiparticles should emerge in the charge transfer
statistics. To see this, we analyze our result for the
transfer statistics in different regimes. We find that
the logarithm of the generating function separates into
two parts, which for very different absolute values of
the electron tunneling amplitudes in the two contacts,
$U_2/U_1\gg1$, become identical with the
cumulant-generating functions of the two separate
contacts. In general, however, each ``single-contact''
term  accounts also for the interference between the
contacts. At low voltages, these single-contact terms
are combined in such a way that the charge transfer
occurs only in electron units, with the low-voltage
asymptotic describing the regular Poisson distribution
of tunneling electrons. In Section 4, we develop
analogous qualitative interpretation of the
large-voltage asymptotics of the Bethe ansatz result for
the generating function. The interpretation is based on
the $m$-state model of the quasiparticle tunneling
described above. Following the method of
Ref.~\onlinecite{nazarov}, the cumulant-generating
function is calculated from the kinetic equation
governing the quasiparticle transitions in the  basis of
the $m$ flux states. Direct comparison of this
generating function with the leading asymptotics of the
Bethe ansatz result reveals their coincidence under a
special choice of $m$ quasiparticle interference phases
which are found as $\phi_l=(\kappa+(m-1)\pi+2\pi l)/m$,
where $l=0,...,m-1$ numbers the flux states of the
interferometer. The  $l$-dependent, statistical, part of
these phases agrees with the expected anyonic statistics
of the quasiparticles, while the common phase is given
by the electron interference phase $\kappa$ with an
additional phase shift $\pi/m$ for even $m$. The
equivalence of the two distributions in the
large-voltage limit proves that the Bethe ansatz
construction we implemented indeed describes the
statistical transmutation of the effective flux through
the MZI. In Section 5, we use the obtained generating
function to find the first three cumulants of the
transferred charge distribution. We study, in
particular, the voltage dependence of the first and the
second cumulants proportional to the average current and
the shot noise, respectively. The ratio of the two, the
Fano factor, is an experimentally observable
\cite{n1,n2} characteristics of the
electron-quasiparticles decomposition. Finally, we
consider the third cumulant, which determines the
asymmetry of the transferred charge distribution around
average. In Section 6, we look at the special case $m=2$
which allows to obtain these cumulants in terms of
elementary functions for arbitrary voltages. The results
of this work are summarized in the Conclusion.

\section{Model of the symmetric Mach-Zehnder
interferometer and its Bethe-ansatz solution}

We start our discussion with the electronic model of MZI
(Fig.~1) formed by two single-mode edges with filling
factors $\nu_l=1/ (2m_l+1)$, $l=0,1$. Electron operator
$\psi_l$ of the edge $l$ is expressed using the standard
bosonization approach \cite{wen} as
\[ \psi_l=(D/2\pi v_l)^{1/2} \xi_l e^{i[\phi_l(x,t)/
\sqrt{\nu_l} +k_lx] } .  \]
Here $\phi_l$ are the two chiral bosonic modes propagating in the same direction (to the right in Fig.~1), which satisfy the usual equal-time commutation relations $[\phi_l(x),\phi_p(0)]=i \pi \mbox{sgn}(x)\,
\delta_{lp}$. The Majorana fermions $\xi_l$ account for the mutual statistics of electrons in different edges, and $D$ is a common
energy cut-off of the edge modes. The Fermi momenta $k_l$ correspond to the average electron density in the edges, while the operators of the density fluctuations are: $\rho_l(x,\tau)= (\sqrt{\nu_l}/2 \pi) \partial_x \phi_l(x, \tau)$.

In the symmetric case of equal times of excitation propagation between the
contacts along the two edges, the two combinations of the bosonic operators $\phi_l$ that enter the electron tunneling terms of the two contacts can be expressed as the values at points $x_{1,2}$ of the same right-propagating
chiral bosonic field:
\begin{equation}
\phi_-(x) ={\sqrt{\nu_1}\phi_0-\sqrt{\nu_0}\phi_1 \over \sqrt{\nu_0 +
\nu_1}} \label{phi-} \, . \end{equation}
The Langrangian describing electron tunneling in the two contacts
can then be written as:
\begin{equation}
{\cal L}_t = \sum_{j=1,2} (DU_j/\pi) \cos [ \lambda
\phi_-(x_j)+ \kappa_j ] \, , \label{e2}
\end{equation}
where $U_j$ and $\kappa_j$ are the absolute values and the phases of the dimensionless tunneling amplitudes. The products of the Majorana fermions $\xi_1 \xi_2$ were omitted from the Lagrangian (\ref{e2}), since they cancel out in each perturbative order due to charge conservation. The phases $\kappa_j$ include contributions from the external magnetic flux $\Phi_{ex}$ and from the average electron numbers $N_{0,1}$ on the two sides of the
interferometer:
\[ \kappa_2 - \kappa_1=2\pi[(\Phi_{ex}/\Phi_0)+ (N_0/\nu_0)-(N_1/\nu_1)]+ \mbox{const} \equiv -\kappa . \]
The factor $\lambda=\sqrt{2m}$ in  the Lagrangian (\ref{e2}) follows
from the normalization of the bosonic field $\phi_-$, which in the absence of tunneling is a free right-propagating chiral field. This field undergoes successive scattering at the two contacts by the tunneling terms of the Lagrangian. The scattering breaks the charge conservation and therefore creates tunneling current. The applied voltage can be introduced at
first as a shift of the incoming field of one of the edges: $\phi_0-\sqrt{\nu_0}Vt$. As one can see from Eqs.~(\ref{phi-}), such a shift translates into the shift of the tunneling field $\phi_--Vt/\lambda$.

The thermodynamic Bethe ansatz solution of the tunneling model was developed \cite{ba} for a single-point tunneling contact with $\lambda^2=2m$ by application of one-particle boundary $S$-matrices \cite{za} to a distribution of the bosonic field excitations (kinks, antikinks, and breathers) introduced through the massless limit of the "bulk" sine-Gordon model. This solution was generalized \cite{usprl} to the two tunneling contacts relevant for the MZI problem by successive application of two boundary $S$-matrices to the  same distribution of the excitations of the model. For the charge transport, only the kink-antikink (and vice verse) transitions are important, and their boundary $S$-matrices are written as
\begin{equation}
{\cal S}^{\pm \pm}_{j,k}={(a k/T_{jB})^{m-1}
e^{i\alpha_k}\over 1+i(a k/T_{jB})^{m-1} }\, , {\cal
S}^{- +}_{j,k} = {e^{i(\alpha_k -\kappa_{j})}\over 1+i(a
k/T_{jB})^{m-1} }\, . \label{e27}
\end{equation}
Here the standard energy scales $T_{jB}$, $j=1,2$, are used to characterize the tunneling strength at the individual contacts, and
\[ a=v {2\sqrt\pi\Gamma(1/[2(1-\nu)]) \over \nu
\Gamma(\nu/[2(1-\nu)])} . \]
The explicit relation between the energy scales $T_{jB}$ and electron tunneling amplitudes is given below.

\section{Cumulant-generating function for the charge transfer distribution}
\label{lnP}

At zero temperature, dynamics of the liquid should be described with only one type of quasiparticles, e.g., kinks, which fill out all available states with the "bulk" distribution $\rho(k)$, with $k$ being the quasiparticle momentum, up to some limiting momentum $A$ defined by the applied voltage. Each momentum-$k$ quasiparticle undergoes successive scattering at the two
tunneling contacts independently of other quasiparticles. The overall scattering process is described by the product of the two boundary $S$-matrices given by Eq.~(\ref{e27}). Our goal is to find the cumulant-generating function $\ln P(\xi )$ of the charge transfer between the two branches of the interferometer, which is defined, as usual, as the logarithm of the Fourier transform of the probability distribution function of the transferred charge. We measure the charge in units of the elementary electron charge by setting $e=1$. The independence of the scattering events of quasiparticles with different momenta implies then that $\ln P(\xi )$ can be found as a sum of logarithm of the generating functions of individual momentum states, and its long-time asymptotics is
\begin{equation}
\ln P(\xi )=t \int_0^{A} dk \rho(k)\ln p(k,\xi) \, .
\label{lnP1}
\end{equation}
The generating function of one state with momentum $k$
\begin{equation}
p(k,\xi)=1+\tau_C(k)(e^{i\xi}-1) \label{p1}
\end{equation}
is defined by the total transition probability $\tau_C(k)$ of
the momentum-$k$ kink into antikink. Taking the product of the scattering matrices of the two contacts to find the total scattering matrix $\hat{\cal S}_2\hat{\cal S}_1$, and using the parametrization of the contact tunneling strengths as $(T_{jB}/a)^2 \equiv \exp\{\theta_j/(m-1)\}$, we can write the  probability $\tau_C(k)$ from the corresponding matrix element of $\hat{\cal S}_2\hat{\cal S}_1$ as
\begin{equation}
\tau_C(k)=|(\hat{\cal S}_2\hat{\cal S}_1)^{-,+}|^2 =B (\tau(\theta_2,k)-\tau(\theta_1,k)) \, .\label{tauC1}
\end{equation}
Here $\tau(\theta_j,k)$ are the transition probabilities in the individual point contacts:
\begin{equation}
\tau(\theta_j,k)=|\hat{\cal S}_j^{-,+}|^2=[1+k^{2(m-1)}e^{-\theta_j}]^{-1} , \label{tau1} \end{equation}
and the factor $B$ characterizes interference between the two contacts:
\begin{equation}
B(T_{jB},\kappa)={|T_{1B}^{m-1}+T_{2B}^{m-1}e^{i\kappa}|^2 \over T_{2B}^{2(m-1)} -T_{1B}^{2(m-1)}}\, . \label{B1}
\end{equation}
Without a loss of generality, we assume below a specific relation between the tunneling strength parameters of the two contacts, $\theta_2 \ge \theta_1$, and write them as $\theta_{1,2}=\bar \theta\mp  \Delta \theta_0$, with $\Delta \theta_0\ge 0$.

The aim of our subsequent derivation is to find the cumulant-generating function $\ln P(\xi )$ in terms of the two generating functions $\ln P_S$ for charge transfer in an individual point contact that was found from the Bethe ansatz solution by Saleur and Weiss \cite{sw}. This derivation
does not need the explicit expressions for $\rho(k)$ and $A$ which can be found in \cite{ba}. Following the approach for one contact, we first relate $\ln P(\xi )$ in Eq.~(\ref{lnP1}) to the effective tunneling current. To
do this, we introduce the generalized tunneling probability $\tau_C(u,k)$:
\begin{equation}
\tau_C(u,k)\equiv[1+(\tau_C^{-1}(k)-1)e^{-u}]^{-1} \, .
\label{tauC2}
\end{equation}
which is the solution of the following differential equation in the new parameter $u$:
\begin{equation}
\partial_u\tau_C(u,k)=(1-\tau_C(u,k))\tau_C(u,k)
\label{tauC3}
\end{equation}
that satisfies the initial condition $\tau_C(u,k)|_{u=0}=\tau_C(k)$. One can extend Eq.~(\ref{p1}) for $p$ to include the parameter $u$ through the substitution $\tau_C(k)\rightarrow \tau_C(u,k)$. Equation (\ref{tauC2}) shows then that the logarithm of $p$ extended this way can be expressed as
\begin{equation}
\ln p=\ln [1+\tau_C(k)(e^{u+z}-1) ]\big{|}_{z=0}^{z=i\xi} \, . \label{der1}
\end{equation}
Calculating the derivatives of Eq.~(\ref{der1}) with respect to $u$ and $\xi$, and  using Eqs.~(\ref{tauC2}), one can see that
\begin{equation}
-i\partial_\xi \ln p=\partial_u \ln p  + \tau_C(u,k) =\tau_C(u+i\xi,k)\, . \label{der2}
\end{equation}
Combining Eq.~(\ref{der2}) with (\ref{lnP1}) in which $p$ is extended to include the parameter $u$, one sees that the cumulant-generating function satisfies the following relation
\begin{equation}
\partial_{i\xi} \ln P(u,\xi )/t=\int_0^{A} dk \rho(k)\tau_C(u+i\xi,k)
\equiv I(u+i\xi,V) \, ,  \label{lnP2}
\end{equation}
which expresses it through the total tunneling current $I(u,V)$ in the two contacts that is defined by the generalized tunneling probability  $\tau_C(u, k)$.

As the next step, one substitutes Eq.~(\ref{tauC1}) into (\ref{tauC2}) and casts the total tunneling probability $\tau_C(u, k)$ into the following form
\begin{equation}
\tau_C(u,k)=B\frac{e^u \sinh\Delta \theta_0}{\sinh\Delta \theta(u)}
\sum \pm \tau(\bar \theta\pm \Delta\theta(u),k)\, ,
\label{tauC4} \end{equation}
where $\Delta \theta(u)$ is defined by the conditions that
\begin{equation}
\cosh\Delta \theta(u)=\cosh\Delta \theta_0 +B(e^u-1)\sinh\Delta
\theta_0 \, ,
\label{theta1} \end{equation}
and $\Delta \theta(u)>0$. Differentiation of Eq.~(\ref{theta1}) shows that the coefficient in Eq.~(\ref{tauC4}) in front of the sum can be written as the derivative of $\Delta \theta (u)$:
\begin{equation}
\partial_u\Delta \theta(u)=B \frac{e^u \sinh\Delta \theta_0}{\sinh\Delta \theta(u)}\, .
\label{theta2} \end{equation}
Using Eqs.~(\ref{tauC4}) and (\ref{theta2}) in the definition of the tunneling current $I(u,V)$ (\ref{lnP2}) we obtain an important relation
\begin{equation}
I(u,V)=\partial_u \Delta \theta(u)\sum_\pm \pm
I_{1/m}(\bar \theta \mp \Delta\theta(u),V)\, , \label{IC}
\end{equation}
which expresses the derivative of the cumulant-generating function (\ref{lnP2}) for the charge transfer in a symmetric interferometer in terms of the tunneling current $I_{1/m}$ in one point contact between the two edges with the  effective filling factor $\nu=1/m$. The current in one contact has been calculated \cite{ba} from the Bethe ansatz solution, and its tunneling conductance
\[G_{1/m}[V/T_B]=G_{1/m}[(V/a) e^{-\theta/[2(m-1)]}]=I_{1/m}(\theta,V) /V\]
is given at zero temperature by a universal scaling function expressed in the form of the low- and high-voltage expansion series. Integrating Eq.~(\ref{IC}) over $u$, and using the result in Eq.~(\ref{lnP2}), we express the generating function $\ln P(\xi)=\ln P(u,\xi)|_{u=0}$ in the following form:
\begin{eqnarray}
\ln P(\xi ) = -Vt\Big\{\int^{\bar \theta- \Delta\theta(i\xi)}_{\theta_{1}} +\int^{\bar \theta+ \Delta\theta(i\xi)}_{\theta_{2}}\Big\} d\theta \nonumber  \\ \cdot \, G_{1/m}[(V/a) e^{-\theta/[2(m-1)]}]
\, .\label{lnP3} \end{eqnarray}
The explicit expansion series for $G_{1/m}$ in Eq.~(\ref{lnP3}) allow integration in each order. The integration transforms the generating function (\ref{lnP3}) into the sum of the two generating functions $\ln P_S$  for charge transfer in individual contacts and gives our key result:
\begin{equation}
\ln P(\xi )=\sum_{j=1,2}\ln P_S \big(V/T_{jB},e^{(-1)^{j} (\Delta\theta_0-\Delta\theta(i\xi))}\big) \, . \label{lnPs1}
\end{equation}
The single-contact generating function $\ln P_S$ is known in terms of the  low- and high-voltage expansion series \cite{sw}:
\begin{eqnarray}
\frac{\ln P_S(s,e^{i\xi} )}{\sigma_0 Vt} = \sum_{n=1}^\infty {c_n(m) \over m n}
s^{2n({1\over \nu}-1)}(e^{i n \xi}-1)\, ,\;\; & s<e^\Delta , \nonumber\\
=i\nu \xi+ \sum_{n=1}^\infty {c_n(\nu)\over n}
s^{2n(\nu-1)}(e^{-i n\nu \xi}-1) \, , \;\; &  s>e^\Delta , \nonumber\\
c_n(\nu)=(-1)^{n+1 }{\Gamma(\nu n+1)\Gamma(3/2)\over
\Gamma(n+1)\Gamma(3/2+(\nu-1)n)}, \;\; & \label{lnPs2}
\end{eqnarray}
where
\[ e^\Delta= (\sqrt{\nu})^{\nu/(1-\nu)}\sqrt{1-\nu}\, , \]
and $\sigma_0$ is the conductance quantum.

Equation (\ref{lnPs1}) representing the total generating
function $\ln P$ as the sum of the two single-contact
generating functions $\ln P_S$ seems to suggest that the
charge transfer in the MZI is divided into two
independent processes associated with the two point
contacts of the MZI. Indeed, one manifestation of this
division is that dependence of each of these processes
on the bias voltage $V$ (through the function $P_S$) is
determined, as in Eq. (\ref{lnPs2}), only by the
characteristics energy scale $T_{jB}$ of the
corresponding junction. Such a division, however, is not
complete. The total generating function $\ln P$ depends
also on the charge dynamics in the interferometer as a
whole, since each individual charge transfer process in
one contact triggers multiple charge transfers involving
interference between both contacts of the
interferometer. Information about this charge dynamics
enters Eq.~(\ref{lnPs1}) through the function $\Delta
\theta(u)$ determined by Eq.~(\ref{theta1}) and
sensitive to both  interferometer contacts.
Nevertheless, the interference can become irrelevant if
the two contacts are strongly asymmetric. For $T_{2B}
\gg T_{1B}$, the generating function $\ln P$ is well
approximated at low voltages by the single-contact $\ln
P_S$ defined by $T_{1B}$, i.e. by the strongest electron
tunneling amplitude $U_1$ -- see Eq.~(\ref{w1}) below.
At large voltages, the generating function $\ln P$ is
approximated by the single-contact $\ln P_S$ defined by
$T_{2B}$, i.e., by the weakest electron tunneling
amplitude $U_2$.

\subsection{Low-voltage behavior of the generating function}

At small voltages, when $V<T_{jB}e^\Delta$ for both $j=1,2$, the low-voltage expansions of the single-contact generating function $\ln P_S$ (\ref{lnPs2}) for both terms in Eq.~(\ref{lnPs1}) can be combined as follows:
\begin{eqnarray}
\ln P(\xi ) = \sigma_0 Vt \sum_{n=1}^\infty {c_n(m) \over m n} \sum_{j=1,2}  \big({V/T_{jB}}\big)^{2n(m-1)} \nonumber \\
\cdot \, \big[{\cosh(n \Delta\theta(i\xi)) \over
\cosh(n \Delta\theta_0)}-1\big] \, .\label{lnP4}
\end{eqnarray}
Since $\cosh(n \Delta\theta(i\xi))$ is a polynomial of $\cosh( \Delta \theta(i\xi))$, and the latter, according to Eq. (\ref{theta1}), is a linear function of $e^{i\xi}$, this expansion of the MZI generating function shows that at low voltages, the charge transfer between the edges of the MZI is quantized in units of electron charge.

More explicitly, using the standard expansion of $\cosh nx$ (see Eq.~1.331.4 in Ref.~\onlinecite{GR}) and the relations that follow from Eqs.~(\ref{B1}) and (\ref{theta1}):
\begin{eqnarray}
\cosh\Delta\theta(i\xi)= \cosh\Delta\theta_0[1+R(e^{i\xi}-1)] \, , \nonumber\\
R\equiv B\tanh\Delta\theta_0=  {|T_{1B}^{m-1}+T_{2B}^{m-1}e^{i\kappa}|^2 \over T_{1B}^{2(m-1)}+T_{2B}^{2(m-1)}} \, ,\label{R} \\
\cosh\Delta\theta_0= {T_{1B}^{2(m-1)}+T_{2B}^{2(m-1)} \over
2(T_{1B}T_{2B})^{(m-1)}} , \nonumber
\end{eqnarray}
we bring Eq.~(\ref{lnP4}) into the following form:
\begin{eqnarray}
\ln P(\xi )= \sigma_0 Vt \sum_{n=1}^\infty {c_n(m)\over m n} \big[\sum_j\big(V/T_{jB} \big)^{2(m-1)}\big]^n \nonumber \\
\cdot \, \Big\{ [1+R(z-1)]^n  +n \sum_{l=1}^{[n/2]} \frac{(-1)^lC^{n-l-1}_{l-1}}{l(2\cosh\Delta\theta_0)^{2l}}
\nonumber \\ \cdot \, \big[ 1+R(z-1) \big]^{n-2l} \Big\}
\Big|^{z=e^{i\xi}}_{z=1} .
\label{lnP5} \end{eqnarray}

Equation (\ref{lnP5}) quantifies our previous conclusions about properties of the MZI charge transfer statistics (\ref{lnPs1}) in the low-voltage regime. In the limit of strongly different contacts, $T_{2B} \gg T_{1B}$,
one finds that $R\to 1$ and $\cosh\Delta\theta_0 \gg 1$, so that the charge transfer statistics (\ref{lnP5}) approaches that of one point contact
\cite{sw} characterized by $T_{1B}$ (i.e., the strongest electron
tunneling amplitude $U_1$) with corrections in $T_{1B}/T_{2B}$ also quantized in the electron charge units. The $n$th term in the expansion of this statistics in powers of bias voltage $V$ corresponds in this case to tunneling of exactly $n$ electrons. By contrast, the $n$th order term of the general MZI transfer statistics (\ref{lnP5}) involves transfer of all numbers of electrons up to $n$. In the lowest order in $V$, the MZI statistics reduces to the Poisson distribution, with the coefficient in front of $(z-1)$ in Eq.~(\ref{lnP5}) is equal to the average electron tunneling current. One can check this starting from Eq.~(\ref{e2}) by direct perturbative calculation \cite{usprl}, if the energy scales $T_{jB}$ are
expressed through the electron tunneling amplitudes $U_j$:
\begin{equation}
T_{jB}= 2 D [\Gamma(m)/U_j] ^{{1/(m- 1)}} . \label{w1}
\end{equation}

\subsection{Large-voltage behavior of the generating function}

At large voltages, $V>T_{jB}e^\Delta$, the combination large-voltage expansions of both single-contact generating functions $\ln P_S$ in Eq.~(\ref{lnPs1}) brings the total MZI generating function into the following form:
\begin{eqnarray}
\ln P(\xi )= \sigma_0Vt \sum_{n=1}^\infty {c_n(1/m)
\over n} \sum_j \big( V/T_{jB} \big)^{2n(1-m)/m}
\nonumber \\  \cdot \,
\big[{\cosh(n\Delta\theta(i\xi)/m)\over\cosh(n
\Delta\theta_0/m) }-1\big] \, . \;\;\; \label{lnP6}
\end{eqnarray}
In terms of the parameters introduced in Eq.~(\ref{R}), this equation can be rewritten as:
\begin{eqnarray}
\ln P(\xi )= \sigma_0Vt \sum_{n=1}^\infty {c_n(1/m)\over n 2^{n/m}} \Big[\sum_j (T_{jB}/V)^{2(m-1)}\Big]^{{n\over m}} \nonumber \\
\cdot\,  \sum_\pm \Big[ 1+R (z-1) \pm \big( [1+R(z-1)]^{2} \nonumber \\ - \cosh^{-2}\Delta\theta_0 \big)^{{1 \over 2}} \Big]^{{n\over m}} \Big|^{z=e^{i\xi}}_{z=1} . \;\;\; \label{lnP7}
\end{eqnarray}
We can see again that in the asymmetric limit, $T_{2B} \gg
T_{1B}$, when  $R\to 1$ and $\cosh\Delta\theta_0 \gg 1$, Eq.~(\ref{lnP7}) for the charge transfer statistics reduces to that of a single point
contact \cite{sw}. The dominant contact is now characterized by the larger quasiparticle tunneling amplitude $W_2$ (i.e., smaller electron tunneling amplitude $U_2$) and corresponding energy scale $T_{2B}$, related to $W_2$ as:
\begin{equation}
T_{jB} =  2 m D [{W_j/\Gamma(1/m)}]^{m/(m-1)}.
\label{w2} \end{equation}

The $n$th order term in the expansion (\ref{lnP7}) of the generating function corresponds to the transfer of the fractional charge
$n/m$ by $n$ quasiparticles. One can see, however, that the MZI transfer statistics (\ref{lnP7}) does not contain in general the terms $e^{i\xi/m} $ that would correspond directly to transfer of individual quasiparticles of charge $1/m$. In particular, the $n=1$ term of the expansion that gives the leading large-voltage contribution to the statistics, can not be interpreted as a Poisson process of tunneling of independent quasiparticles, in contrast to the leading low-voltage term that did represent Poisson process of individual electron tunneling events. The reason for this is the $m$-state dynamics of the effective flux through the interferometer associated with the quasiparticle tunneling, which introduces correlations in the tunneling process. These  correlations can be most easily understood in the description of the quasiparticle tunneling based on kinetic equation. Such an equation is discussed in the next Section.

\section{Kinetic equation for the large-voltage charge transfer}

To derive kinetic equation that reproduces the
large-voltage asymptotics of the generating function
(\ref{lnP6}), and therefore provides a simple physical
picture of the dynamics of quasiparticle tunneling, we
start by rewriting this asymptotics in terms of the
quasiparticle tunneling amplitudes $W_j$. From the last
relation in Eq.~(\ref{R}), we have
\begin{equation} \cosh (\Delta \theta_0 /m) = \frac{1}{2} \Big[ \big(\frac{T_{1B}}{T_{2B}} \big)^{\frac{m-1}{m}} + \big(\frac{T_{2B}}{T_{1B}}\big)^{\frac{m-1}{m}} \Big] . \label{der5}
\end{equation}
Using this equation to transform the leading, $n=1$, term in Eq.~(\ref{lnP6}), and replacing the energy scales $T_{jB}$ by the quasiparticle amplitudes $W_j$ with the help of Eq.~(\ref{w2}), we have:
\begin{equation}
\ln P(z)=t K(V)\big[ 2W_1W_2\cosh \big(\frac{\Delta\theta(z)}{m}\big) -\sum_j W^2_j\big]  \, ,
\label{lnP8} \end{equation}
where $K(V)=\sigma_0V(2mD/V)^{2(m-1)/m} c_1(1/m)/\Gamma^2(1/m)$.

Kinetic equation describing the quasiparticle tunneling
can be written down based on the following
considerations. The general picture of the quasiparticle
dynamics in the MZI discussed in the Introduction
implies that the quasiparticles create statistical
contribution to the effective flux through the
interferometer. Because of this statistical
contribution, the MZI can be found in $m$ separate states
which differ by the effective flux, with each tunneling
quasiparticle changing successively the state $l$ into $l-1$
modulo $m$. Since the total rates of the quasiparticle
tunneling in the MZI depend on the effective flux, such
dynamics of flux introduces correlations into
quasiparticle transitions, separating naturally the
processes of successive quasiparticle tunneling events
into the groups of $m$ transitions. As usual, to cast
the kinetic equation governing this flux dynamics into
the form appropriate for the calculation of the
cumulant-generating function, we multiply transition
probabilities by a factor $z^{1/m}=e^{i\xi/m}$ that
keeps track of the transferred charge. Then, we
introduce the probabilities $d_{l,n}(t)$ that at time
$t$ the MZI is in the state $l$ and $n$ quasiparticles
have been transferred through it. Combining these
probabilities into an $m$-dimensional vector
$Q_l(z,t)=\sum_n d_{l,n}(t) z^{n/m}$, one can write the
kinetic equation in the following matrix form:
\begin{equation}
\partial_t Q_l(z,t)=\sum_{l'} M(z)_{l,l'} Q_{l'}(z,t)
\, .\label{kineq}
\end{equation}

According to the qualitative picture of quasiparticle tunneling discussed above, the transition matrix has a simple form, with the only non-vanishing elements are those on the main diagonal, $l=l'$, and those with $l=l'-1$:
\begin{equation}
M(z)_{l,l'}=-\gamma_l\delta_{l,l'}+\gamma_{l'}z^{1/m}\delta_{l,l'-1}
\, .\label{M}
\end{equation}
Here the Kronecker symbol $\delta_{l,l'}$ is defined modulo $m$. The leading large-time asymtotics of the generating function for the probability distribution evolving according to the kinetic equation (\ref{kineq}) is (see, e.g., \cite{nazarov}): $\ln P(z)=t\Lambda$, where $\Lambda$ is the maximum eigenvalue of the transition matrix (\ref{M}). The structure of this matrix shows directly that the characteristic equation $\det (M- \Lambda) =0$ has the form:
\begin{equation}
\prod_{l=0}^{m-1} (\gamma_l+\Lambda)-z\prod_{l=0}^{m-1} \gamma_l=0\,
. \label{Lambda1}
\end{equation}
The maximum eigenvalue $\Lambda$ is the solution of this
equation which goes to zero when $z \to 1$, since all
other eigenvalues of the matrix $M(z=1)$ are negative.

Before trying to establish the general relation between
the generating function obtained from this equation and
the generating function (\ref{lnP8}), we consider the
simple case $m=2$. In this case, Eq.~(\ref{lnP8}) for
the large-voltage asymptotics of the generating function
(\ref{lnP6}), can be simplified further. First, we have
from Eq.~(\ref{R}):
\[ \cosh (\frac{\Delta \theta (z)}{2}) = \Big[\cosh^2(\frac{\Delta \theta_0}{2})+\frac{R}{2} (z-1) \cosh \Delta \theta_0 \Big]^{1/2} . \]
Using this relation, Eq.~(\ref{R}), and Eq.~(\ref{der5}) with $m=2$, we transform Eq.~(\ref{lnP8}) into:
\begin{eqnarray}
\ln P(z)=t K(V)\Big\{ \Big[ (\sum_j
W^2_j)^2+|W_1^2+W_2^2e^{i\kappa}|^2 \nonumber \\ \cdot
\, (z-1)\Big]^{1/2}\!\!\! -\sum_j W^2_j\Big\} .
\label{lnP9}
\end{eqnarray}
On the other hand, in the kinetic-equation approach, Eq.~(\ref{Lambda1})
 be solved readily for $m=2$ giving the following expression for $\Lambda$:
\begin{equation}
\Lambda=\Big(\big[\big(\sum_j \gamma_j\big )^2 +4\gamma_0 \gamma_1 (z-1) \big]^{1/2} -\sum_j \gamma_j \Big)/2 \, .
\label{Lambda2}\end{equation}

This equation describes the generating function for the statistics of any transfer process consisting of the two steps with the rates $\gamma_{0,1}$, e.g., incoherent charge transfer through a resonant level \cite{nazarov}.
Comparison of Eqs.~(\ref{Lambda2}) and (\ref{lnP9}) shows that the generating function (\ref{Lambda2}) obtained from the kinetic equation reproduces the large-voltage asymptotics (\ref{lnP9}) of the generating function  (\ref{lnP6}), if the tunneling rates are taken as
\begin{equation}
\gamma_{0,1}=K(V)|W_1+W_2e^{i\phi_{0,1}}|^2, \;\;
\phi_l=(\kappa+\pi)/2+\pi l \, .
\label{emp1}  \end{equation}
Equation (\ref{emp1}) for the tunneling rates agrees with the physical picture of quasiparticle tunneling discussed above. Statistical contribution to the effective flux though the MZI means that tunneling of each quasiparticle changes the phase between the interferometer contacts by $2\pi/m=\pi$ in agreement with the quasiparticle anyonic exchange statistics. Equation (\ref{emp1}) also shows that the quasiparticles see a phase shift $\pi/2$ in addition to the phase $\kappa/2$ induced by the external magnetic flux. The origin of this phase shift is discussed below.

Following this logic, we look for solution of Eq.~(\ref{Lambda1}) in the case of arbitrary $m$ taking the tunneling rates $\gamma_l$ as:
\begin{equation}
\gamma_l=K(V)|W_1+W_2e^{i\phi_l}|^2, \;\;\; \phi_l=\phi+2\pi l/m \, ,
\label{emp2}  \end{equation}
with some unknown $\phi$. With these tunneling rates, Eq.~(\ref{Lambda1}) reads
\begin{equation}
\prod_{l=0}^{m-1} (\cos \phi_l+{\lambda+W_1^2+W_2^2
\over 2 W_1W_2})=z\prod_{l=0}^{m-1} {\gamma_l\over 2K(V)
W_1W_2} \, ,\label{Lambda3}
\end{equation}
where $\lambda \equiv \Lambda/K(V)$. To further transform Eq.~(\ref{Lambda3}), we use the basic identity:
\begin{equation}
x^m-1=\prod_{l=0}^{m-1}(x-e^{i2\pi l/m})\, , \label{identity}
\end{equation}
and the two identities that follow directly from it:
\begin{eqnarray}
\prod_{l=0}^{m-1}2 \cos{\phi_l\over 2}= 2\cos({\phi
\over 2}m + \pi{m-1 \over 2}) \, , \nonumber \\
2^{m-1} \prod_{l=0}^{m-1}[\cos\phi_l+ \cosh
(\frac{\Delta\theta}{m})] =\cosh\Delta\theta  \label{iden} \\
- \, (-)^m \cos(m\phi) \, .\nonumber
\end{eqnarray}
The first one is obtained essentially by taking $x=-e^{i\phi}$ in Eq.~(\ref{identity}), while the second one follows from the first if the sum of the cosines is transformed into their product. By direct comparison of
the second identity in (\ref{iden}) with Eq.~(\ref{Lambda3}) we see that $\Lambda$ as defined by $\ln P(z)/t$ in Eq.~(\ref{lnP8}) indeed solves Eq.~(\ref{Lambda1}) if
\begin{equation}
\cosh\Delta\theta=(-)^m\cos(m\phi)+{z\over 2} \prod_{l=0}^{m-1} {\gamma_l\over  K(V) W_1 W_2}\, .
\label{RR} \end{equation}
Making use of Eq.~(\ref{identity}) one more time, we calculate the product on the right-hand-side of Eq.~(\ref{RR}):
\begin{equation}
\prod_{l=0}^{m-1} {\gamma_l\over K(V) W_1 W_2}= \frac{|W_2^me^{im(\phi-\pi)} -W_1^m|^2}{W_1^m W_2^m}\,
. \label{productgammas}
\end{equation}

One can see that with this expression for the product, Eq.~(\ref{RR}) precisely coincides with the definition of $\cosh\Delta\theta(z)$ by Eq.~(\ref{R}). Indeed, combining all three relations in Eq.~(\ref{R}) one can express $\cosh\Delta\theta(z)$ as
\begin{equation}
\cosh\Delta \theta(z)=-\cos\kappa +{|T_{1B}^{m-1}+T_{2B}^{m-1} e^{i\kappa}|^2 \over 2 (T_{2B}T_{1B})^{(m-1)}}z \, .
\label{theta3} \end{equation}
Replacing $T_{jB}$ in Eq.~(\ref{theta3}) with the amplitudes $W_j$ through
Eq.~(\ref{w2}), we see that  Eq.~(\ref{theta3}) precisely coincides with Eqs.~(\ref{RR}) and (\ref{productgammas}), if $\cos \kappa= (-)^{(m+1)} \cos(m\phi)$, i.e., if the phase $\phi$ is chosen to satisfy the condition
\begin{equation}
m \phi=\kappa +(m-1)\pi \, . \label{fi}
\end{equation}
The deviation of the phase $\phi$ from $\kappa/m$ in this equation is important only for even $m$. In this case, it produces the shift $(m-1)\pi/m$ of the interference phase in the quasiparticle tunneling rates from the value induced by the external magnetic field. This shift coincides with the phase acquired by one of the two Klein factors of the MZI quasiparticle tunneling action of the MZI \cite{usprb} in the flux-diagonal
representation due to the $m$-power periodicity condition for the
Klein factors, which corresponds physically to the requirement of the proper exchange statistics between electrons and quasiparticles. This phase shift
ensures, actually, that there is no shift in the interference pattern of the  tunnel current in the interferometer (see the final results below) between the regimes of electron and quasiparticle tunneling.

\section{Cumulants of the charge transfer distribution}

So far, we have established the interpretation of the
low- and high-voltage asymptotic behavior of the charge
transfer statistics in terms of, respectively,
tunneling of individual electrons and quasiparticles. In
this Section, we calculate the charge transfer cumulants
in these two limits, with the emphasis on the
quasiparticle limit which exhibits the non-trivial behavior
of the cumulants. We also will use the generating
function found in Sec.~\ref{lnP} for arbitrary voltages
to calculate the full voltage dependence of the
cumulants, and to study the crossover between the two
asymptotic regimes of electron and quasiparticle
tunneling. The cumulants of the charge $N(t)$
transferred through the interferometer during a large
time interval $t$ can be found from the
cumulant-generating function (\ref{lnP2}) by the
standard relation:
\begin{equation}
\langle N^j(t)\rangle_c/t=\partial_u^j \ln P|_{u=0}/t  =\partial_u^{j-1}I(u,V)|_{u=0}\, ,
\label{cum}
\end{equation}
where the current $I(u,V)$ is given by Eq.~(\ref{IC}).

The first cumulant gives the average tunneling current: $I(V) \equiv I(0,V)= \langle N(t)\rangle/t$. At arbitrary voltages, the average current was calculated before in \cite{usprb}. Its large-$V$ asymptotics can also be obtained directly from Eqs.~(\ref{lnP7}) and (\ref{w2}):
\begin{equation}
I(V)= \frac{K(V)}{m}B(W^2_2-W^2_1) = {1 \over \sum_{l} \gamma_{l}^{-1}} \, , \label{I1} \end{equation}
where the interference factor $B$ (\ref{B1}) can be expressed in term of  the quasiparticle amplitudes $W_{1,2}$ as
\begin{equation}
B(W_j,\kappa)= {|W_{1}^{m}+W_{2}^{m}e^{i\kappa}|^2 \over
W_{2}^{2m}-W_{1}^{2m}} \, . \label{B2} \end{equation}
The second equality in Eq.~(\ref{I1}) provides direct
interpretation of the asymptotics of the average current
in terms of the quasiparticles transitions \cite{mz4}.
It can be proven formally by making use of the identity
\begin{equation}
\sum_{l} \gamma_{l}^{-1}= \frac{W_1\partial_{W_1}-W_2
\partial_{W_2}}{2K(V)(W_1^2-W_2^2)} \ln[ \prod_{l} \gamma_{l}]
\, , \label{sum1}
\end{equation}
which can be obtained by directly differentiating individual rates $\gamma_l$ in this equation. On the other hand, differentiating the product of all $\gamma_l$s together, as given by Eq.~(\ref{productgammas}), and using Eq.~(\ref{fi}), we obtained the second equality in (\ref{I1}). This equality agrees naturally with the simple solution of the quasiparticle kinetic equation, which gives the average tunneling rate as inverse of the average tunneling times in the different states of the interferometer.

The second cumulant $\langle N^2(t) \rangle$ defines the spectral density of the current fluctuations at zero frequency $S_I(0)=\langle N^2(t) \rangle_c /t$, which at zero temperature reflects the shot noise associated with the charge transfer processes. From Eq.~(\ref{IC}), one can write the first derivative of the current as
\begin{eqnarray}
\partial_u I(u,V)=I(u,V) \times \partial_u\ln[\partial_u\Delta \theta(u)] \nonumber \\ -(\partial_u\Delta\theta(u))^2 \sum_\pm \partial_{\bar \theta } I_{1/m}(\bar \theta\mp \Delta\theta(u),V)\, .
\label{dIC} \end{eqnarray}
Substituting this formula into Eq.~(\ref{cum}) and calculating the derivatives of $\Delta\theta(u) $ (\ref{theta2}) for $u \rightarrow 0$
we obtain expression for the spectral density of current:
\begin{equation}
S_I(0) =(1-B\coth \Delta \theta_0)I -B^2\sum_{j=1,2}
\partial_\theta I_{1/m}(\theta_j,V)\, . \label{I2}
\end{equation}
It is convenient to characterize the short noise represented by this spectral density through the Fano factor $F$ defined as $F=S_I(0)/I$. In the case of MZI, the Fano factor reflects both the charge and statistics of the tunneling excitations and illustrates the transition between the electron and quasiparticle regimes. In the low-voltage limit, $F=1$ as a result of the regular Poisson process of electron tunneling. To find the Fano factor in the quasiparticle, large-voltage, limit, we start with Eq.~(\ref{I2}) which gives the following general expression for $F$:
\begin{eqnarray}
F=1-B\Big\{\coth \Delta \theta_0 - \sum_j \partial_\theta I_{1/m}(\theta_j, V) \nonumber \\ \cdot \big[ \sum_j (-1)^j I_{1/m}(\theta_j,V) \big]^{-1}  \Big\} \, . \;\;\;
\label{FF} \end{eqnarray}
Using Eq.~(\ref{R}), the fact that in the large-voltage limit only one quasiparticle tunneling term $\propto W^2$ can be kept in the current $I_{1/m}$, cf. Eq.~(\ref{I1}), and that with the parametrization of the energy scales $T_{jB}$ with $\theta$ introduced above, $W^2\propto e^{\theta/m}$, we get from Eq.~(\ref{FF}):
\begin{equation}
F=1-B\Big\{{W_{2}^{2m}+W_{1}^{2m} \over W_{2}^{2m}-W_{1}^{2m}} -{1\over m}{W_{2}^{2}+W_{1}^{2} \over W_{2}^{2}-W_{1}^{2}}\Big\} .
\label{F} \end{equation}

The Fano factor (\ref{F}) corresponds to the dynamics of quasiparticle tunneling as described by the kinetic equation (\ref{kineq}). This can be seen by following the steps similar to that taken above for the average current. Applying the differential operator from Eq.~(\ref{sum1}) to individual terms in the sum of the inverse tunneling rates $\gamma_l$, one obtains directly the following identity:
\[ - \frac{W_1\partial_{W_1}-W_2 \partial_{W_2}}{2K(V)(W_1^2-W_2^2)}
\sum_l\gamma_l^{-1} = \sum_l \gamma_l^{-2} \, . \]
On the other hand, replacing the sum of inverse $\gamma_l$s in this equation  with the corresponding expression from Eq.~(\ref{I1}):
\[ \sum_l \gamma_l^{-1} = m/[B K(V)(W^2_2-W^2_1)] \, , \]
and performing differentiation, we see that the large-voltage asymptotics (\ref{F}) of the Fano factor can be written in terms of the tunneling rates $\gamma_l$ as
\[F=\sum_l \gamma_l^{-2}/(\sum_l \gamma_l^{-1})^2 \, , \]
This result agrees with the calculation \cite{feldman2} based directly on the  kinetic equation. Because of the complex nature of the quasiparticle tunneling dynamics characterized by $m$ different tunneling rates $\gamma_l$, $F$ is not equal simply to the quasiparticle charge $1/m$ but
varies as a function of parameters, e.g. the interference phase $\kappa$,   between $1/m$ and $1$.

\begin{figure}[htb]
\setlength{\unitlength}{1.0in}
\begin{picture}(3.,2.6)
\put(-0.2,-0.15){\epsfxsize=3.4in \epsfbox{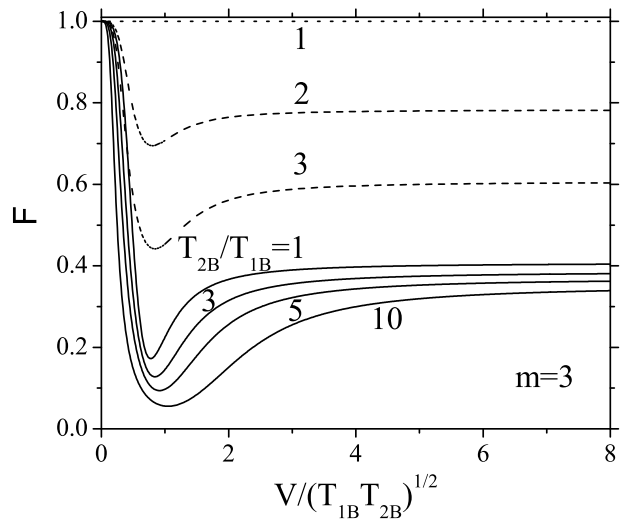}}
\end{picture}
\caption{The zero-temperature Fano factor $F$ of the tunnel current
in the Mach-Zehnder interferometer formed by two $\nu=1/3$ edges,
i.e. for $m=3$, as a function of the bias voltage $V$ for different
degrees of asymmetry of the tunneling strength of the two
contacts characterized by the $T_{1B}/T_{2B}$ ratio. The solid curves
corresponds to the case of complete constructive interference,
$\kappa=0$; for the dashed curves, $\kappa=\pi$. In the latter case,
$F=1$ identically for identical contacts, $T_{1B}/T_{2B}=1$. The curves illustrate the transition between the electron regime $F=1$ at small voltages to the quasiparticle $m$-state tunneling dynamics at large voltages. The transition region is characterized by the Fano factor $F$ reaching the minimum below the quasiparticle minimum $1/m=1/3$.}
\end{figure}

At arbitrary bias voltage $V$, the Fano factor $F$ should be plotted
numerically. Figure 2 shows $F$ in the case $m=3$ which corresponds,
e.g., to tunneling between the two $\nu=1/3$ edges. The curves are
shown for different degrees of asymmetry between the two contacts of
the interferometer and two values of interference phase, maximum
constructive interference, $\kappa=0$, and complete destructive
interference, $\kappa=\pi$. The range of variation of $F$ with the
interference phase $\kappa$ decreases with increasing junction
asymmetry. In general, the curves show the transition between
electron tunneling with $F=1$ at small voltages $V$ to quasiparticle
tunneling at large voltages. In the quasiparticle regime, $F$ is
still can be significantly different from $1/3$ because of the
non-trivial three-state flux dynamics of the MZI. In particular, for identical junctions,  the three total quasiparticle tunneling rates (\ref{emp2}) satisfy the relation: $\gamma_l \propto \cos^2 (\phi_l/2) $.  Taking into account Eq.~(\ref{fi}) for the non-statistical contribution to the interference phase, one can see that under the condition of destructive interference, $\kappa \simeq \pi$, the tunneling rate in one of the flux  states of the interferometer, $l=0$, is much smaller than the rates in the two other states. This mean that on the relevant large time scale set by the smallest rate, the three quasiparticles transition that transfer interferometer from state $l=0$ back to itself happen almost simultaneously, so that the three quasiparticle charges $1/3$ are effectively transferred together, restoring $F$ back to 1.

Finally, we study the third charge transfer cumulant that characterizes the asymmetry around average of the transferred charge distribution, and has  been measured experimentally for electron tunneling in metallic tunnel  junctions -- see, e.g., \cite{3d1,3d2}, and in quantum point contacts \cite{3d3}. As for the other cumulants, the large-time asympotic of the third cumulant is linear in time, and it can be characterized by the coefficient $C_3\equiv \langle N^3\rangle_c /t$. To calculate this coefficient, we first find the second derivative of the tunnel current from Eq.~(\ref{dIC}):
\begin{eqnarray}
\partial^2_u I(u,V)=\big( \partial^2_u\ln\partial_u \Delta\theta(u)+ [\partial_u\ln\partial_u \Delta\theta(u)]^2 \big)I - \nonumber \\ \sum_\pm \big[ 3\partial_u\Delta\theta  \partial^2_u\Delta\theta \partial_{\bar \theta} \mp (\partial_u\Delta\theta)^3 \partial^2_{\bar \theta }\big]
I_{1/m}(\bar \theta\mp \Delta\theta,V). \;\;\;
\label{ddIC} \end{eqnarray}
Derivatives of $\Delta\theta(u)$ here can be found from  Eq.~(\ref{theta2}). In particular, the coefficient in front of $I$ in Eq.~(\ref{ddIC}) can be expressed as:
\begin{eqnarray}
\partial^2_u\ln\partial_u\Delta \theta(u) +[\partial_u\ln\partial_u\Delta \theta(u)]^2=1- 3\coth \Delta \theta(u)\nonumber \\
\times \partial_u \Delta\theta(u) + [3\coth^2 \Delta \theta(u)-1] (\partial_u\Delta\theta(u))^2\, .
\label{coeff} \end{eqnarray}
Substitution of Eqs.~(\ref{ddIC}), (\ref{coeff}), and (\ref{theta2}) into Eq.~(\ref{cum}) gives us the coefficient $C_3$ of the third cumulant:
\begin{eqnarray}
C_3 =\big[ 1-3 B\coth \Delta \theta_0+B^2(3\coth^2 \Delta \theta_0-1)\big] I \nonumber \\  - \sum_{j=1,2} \big[3 B^2 (1-B\coth \Delta \theta_0)
\partial_\theta I_{1/m} (\theta_j,V) \;\;\; \label{C3-2} \\
+ (-)^jB^3\partial^2_\theta I_{1/m}(\theta_j,V) \big] . \nonumber  \end{eqnarray}

The ratio $F_3=C_3/I$ has also been suggested \cite{c3} as a possible  alternative to the Fano factor to characterize the charge of the tunneling particles. Indeed, in a Poisson process,  $F_3$ is equal to the Fano factor multiplied by the tunneling charge. Therefore, in the case of MZI,  $F_3$ (\ref{C3-2}) reduces to 1 in the low-voltage limit, as a result of the regular Poisson electron tunneling. In the quasiparticle, large-voltage, limit, repeating the calculation similar to that leading to Eq.~(\ref{FF}), we can relate both factors as follows:
\begin{eqnarray}
F_3=3F-2+ B^2\Big\{2+ \frac{1}{m^2}+{12W_{2}^{2m}W_{1}^{2m}
\over (W_{2}^{2m}-W_{1}^{2m})^2} \nonumber \\
-\frac{3}{m}{(W_{2}^{2}+W_{1}^{2}) \over
(W_{2}^{2}-W_{1}^{2})}{(W_{2}^{2m}+W_{1}^{2m}) \over
(W_{2}^{2m}-W_{1}^{2m})} \Big\} \, . \label{barF}
\end{eqnarray}

\section{Charge transfer statistics for $m=2$.}

For general $m$, the results for the cumulants of the charge transfer statistics in the MZI discussed above can not be presented in a finite analytical form for arbitrary bias voltages. The situation is simpler for $m=2$, when the kink-quasiparticles of the "bulk" sine-Gordon model that provide the basis for the Bethe-ansatz solution of the MZI transport are the regular fermions (though carrying charge $1/2$), and their distribution $\rho(k)$ in Eq.~(\ref{lnP1}) is the Fermi-Dirac step-function. In practice, the $m=2$ regime should take place in the MZI formed by the two edges of different Quantum Hall liquids, with filling factors $\nu=1/3$ and $\nu=1$. Individual tunneling contacts of this type have been realized experimentally \cite{expm2}. Using the  Fermi-Dirac property of the distribution function $\rho(k)$  for $m=2$, one can obtain the cumulant generating function $\ln P(\xi)$ either by integration in Eq.~(\ref{lnP1}), or equivalently, by direct substitution into Eq.~(\ref{lnP3}) of the known single-contact tunnel conductance $G_{1/2}$, which can be expressed simply as $G_{1/2}(s)=\sigma_0 [1-\arctan(2s)/(2s)]/2$. The resulting  generating function is:
\begin{eqnarray}
\ln P(\xi )=\sigma_0 V t \sum_{j=1,2} \Big( y_j(u) \arctan[ 1/y_j(u)]
\nonumber \\ + (1/2) \ln[1+y^2_j(u)] \Big) \Big|^{u=i\xi}_{u=0} \, ,
\label{m2} \end{eqnarray}
where $y_j(u)\equiv y_j(0)\exp\{(-1)^j (\Delta \theta(u)-\Delta
\theta_0)/2 \}$, and $y_j(0)\equiv T_{jB}/(2V)$.

This generating function can be combined with Eq.~(\ref{cum}) to calculate the cumulants of the transferred change distribution for $m=2$ using the same steps as in the previous section. Alternatively, one can substitute Eq.~(\ref{lnP1}) into (\ref{cum}) and perform the integration directly. Below we briefly discuss the first three cumulants obtained in this way. The first cumulant gives the average tunneling current in the MZI as \cite{usprb}:
\begin{equation}
I= \sigma_0 B ( \Gamma_2 I_2 - \Gamma_1 I_1 ) \, ,
\label{Im2} \end{equation}
where $I_j\equiv \arctan (V/2\Gamma_j)$, and $\Gamma_j \equiv T_{jB}/4 =D W_j^2/\pi$ are the characteristic quasiparticle tunneling rates in separate contacts. Using the fact that $K(V)=\sigma_0 D$ for $m=2$, one can see explicitly that the current agrees at large voltages with Eq.~(\ref{I1}) that follows from the quasiparticle kinetic equation with the tunneling rates (\ref{emp1}):
\begin{equation}
I = \frac{\pi \sigma_0}{2} {|\Gamma_1+\Gamma_2 e^{i\kappa} |^2 \over
\Gamma_1 +\Gamma_2}= \frac{\gamma_0 \gamma_1}{\gamma_0+ \gamma_1} \, . \label{IIm2}
\end{equation}
It is interesting to note that this agreement relies strongly on the shift of the interference phase (\ref{fi}) from the externally-induced phase $\kappa$.

\begin{figure}[htb]
\setlength{\unitlength}{1.0in}
\begin{picture}(3.,2.6)
\put(-0.2,-0.15){\epsfxsize=3.4in \epsfbox{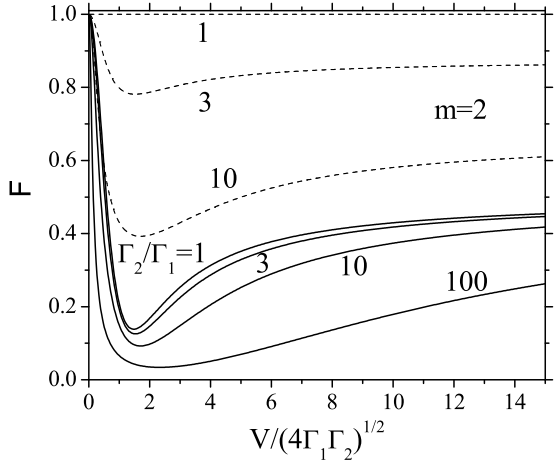}}
\end{picture}
\caption{The zero-temperature Fano factor $F$ of the tunnel current for $m=2$, i.e., in the Mach-Zehnder interferometer formed by edges with $\nu=1/3$ and $\nu=1$, \protect as given by Eq.~(\ref{Fm2}). The solid curves corresponds to the case of complete constructive interference, $\kappa=0$; for the dashed curves, $\kappa=\pi$. In the latter case, $F=1$ identically for identical contacts. In general, the transition from the electron regime $F=1$ at small voltages to the $2$-state tunneling dynamics of quasiparticles at large voltages is characterized by the Fano factor $F$ reaching the minimum in the crossover region. \label{Fonehalf} } \end{figure}

The second cumulant gives the following expression for the Fano factor at  arbitrary voltages, including the transition region between electron and quasiparticle tunneling:
\begin{eqnarray}
F=1-\frac{|\Gamma_1+\Gamma_2e^{i\kappa}|^2}{\Gamma_1I_1-\Gamma_2I_2}
\Big\{ \frac{2\Gamma_1\Gamma_2(\Gamma_2I_1-\Gamma_1 I_2)}{
(\Gamma_1^2 -\Gamma_2^2)^2}+ \nonumber \\
\frac{1}{2( \Gamma_1^2 -\Gamma_2^2)}\sum_{j=1,2} \Gamma_j \big[I_j+
\frac{V/2\Gamma_j}{1+(V/2\Gamma_j)^2}\big] \Big\} \, . \label{Fm2}
\end{eqnarray}
This equation is plotted in Fig.~(\ref{Fonehalf}) and describes the transition from $F=1$ for electron tunneling at small voltages to
\[ F=1- \frac{1}{2} \frac{|\Gamma_1+\Gamma_2e^{i\kappa}|^2}{(
\Gamma_1+\Gamma_2)^2} \]
for quasiparticle tunneling at large voltages. One can see that the
quasiparticle charge $e/2$ manifests itself most clearly for
$\kappa=0$, when the total quasiparticle tunneling rates (\ref{emp1}) coincide, $\gamma_1=\gamma_2$, regardless of the relation between the
individual rates $\Gamma_j$. Similarly to the case $m=3$ illustrated in Fig.~2, the Fano factor reduces to electron value 1 even in the quasiparticle regime, if $\Gamma_1\simeq \Gamma_2$ and $\kappa=\pi$. In this case, one of the total quasipartical tunneling rates $\gamma$ is much smaller than the other, so on the relevant large time scale set by the smaller rate the quasiparticles effectively tunnel together, restoring $F$ to 1.

\begin{figure}[htb]
\setlength{\unitlength}{1.0in}
\begin{picture}(3.,2.6)
\put(-0.2,-0.15){\epsfxsize=3.4in \epsfbox{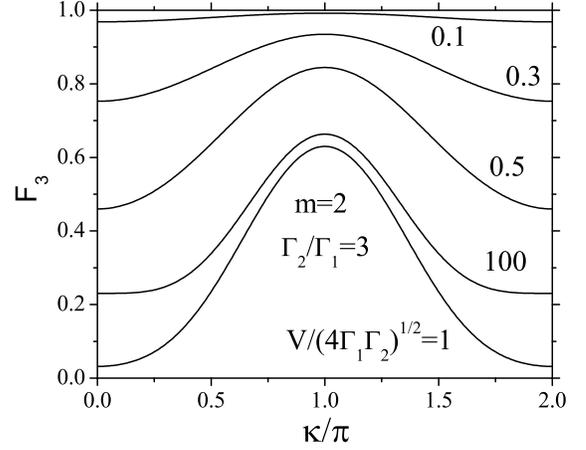}}
\end{picture}
\caption{Alternative ``Fano factor'' $F_3=C_3/I$ \protect (\ref{last}) related to the third cumulant of the tunnel current noise in the Mach-Zehnder interferometer with $m=2$ and zero temperature, as a function of the interferometer phase $\kappa$. The curves are plotted for several bias voltages $V$ between the interferometer edges, and illustrate the transition between the electron and quasiparticles tunneling with increasing  voltage. The transition is characterized by the non-monotonous change of $F_3$, which reaches minimum at the intermediate voltages. \label{F3onehalf} } \end{figure}

Calculating the third cumulant, we find the following expression for the ``alternative'' Fano factor:
\begin{eqnarray}
F_3=3F-2+ \frac{|\Gamma_1+\Gamma_2e^{i\kappa}|^4}{4 ( \Gamma_1I_1- \Gamma_2I_2)( \Gamma_1^2 -\Gamma_2^2)^2 }  \sum_{j=1,2} \Big\{ (-1)^{j+1}   \nonumber \\ \cdot \,\frac{3\Gamma_jI_j (\Gamma_j^4 +10\Gamma_j^2\Gamma_{j'}^2+5\Gamma_{j'}^4) }{( \Gamma_1^2 -\Gamma_2^2)^2} + \frac{3 V }{2( \Gamma_1^2 -\Gamma_2^2)} \;\;\;\;  \label{last} \\
\cdot \, \frac{\Gamma_j^2+3\Gamma_{j'}^2 }{1+(V/2\Gamma_j)^2} +(-1)^{j+1}\frac{V}{[1+(V/2\Gamma_j)^2]^2} \Big\} \nonumber \, , \end{eqnarray}
where $j'$ is defined by $j,j'=1,2$, $j'\neq j$. Equation (\ref{last}) is plotted in Fig.~\ref{F3onehalf}, which shows $F_3$ as a function of the interference phase $\kappa$ at several voltages. Voltage dependence of $F_3$ is qualitatively very similar to that of the Fano factor shown in Fig.~(\ref{Fonehalf}): it approaches 1 at small voltages, in agreement with the underlying Poisson tunneling process of electrons. At large voltages, Eq.~(\ref{last}) reduces to the following form
\[ F_3 =3F-2+  \frac{3}{4} \frac{|\Gamma_1+\Gamma_2e^{i\kappa}|^4}{(
\Gamma_1+\Gamma_2)^4} \, . \]
This expression can be understood in terms of the same two-state tunneling dynamics of the quasiparticles that was discussed above. Also similarly to the Fano factor, the voltage dependence of $F_3$ is non-monotonic, with a minimum  between the regimes of electron and quasiparticle tunneling. The main qualitative difference between the noise-related Fano factor and its   third-cumulant alternative is that the minimum of $F_3$ can become negative for some values of parameters (regime not shown in  Fig.~\ref{F3onehalf}).

\section{Conclusion}

Starting from the exact solution of the tunneling model
of symmetric Mach-Zender interferometer in the FQHE
regime, we have calculated the statistics of the charge
transfer between interferometer edges. The obtained
statistics shows the transition from electron tunneling
at low voltages to tunneling of anyonic quasiparticles
of the fractional charge $e/m$ and statistical angle
$\pi/m$ at large voltages. Deep in the electron
tunneling regime, the dynamics of charge transfer is
represented by the standard Poisson process. Dynamics of
quasiparticle tunneling is more complicated and reflects
the existence of $m$ effective flux states of the
interferometer. The interference phase between the
quasiparticle tunneling amplitudes in two contacts of
the interferometer contains a contribution from the
quasiparticle exchange statistics, making the
quasiparticle tunneling rates in different
interferometer states different. In general, the
transition from electron to quasiparticle tunneling is
reflected in the Fano factor $F$ or its third-cumulant
alternative $F_3$, which both reach minima in the
transition region. However, in the regime close to
complete destructing interference (interferometer phase
$\kappa=\pi$ and equal tunneling strength in the two
contacts), both $F$ and $F_3$ have electron value 1 for
all voltages.

V.P. acknowledges support of the ESF Science Program INSTANS, and
the grant PTDC/FIS/64926/2006.


\begin{thebibliography}{99}


\bibitem{mz1} Y. Ji,
Nature {\bf 422}, 415 (2003); I.
Neder, M. Heiblum, Y. Levinson, D. Mahalu, and V.
Umansky, Phys.\ Rev.\ Lett. {\bf 96}, 016804 (2006).

\bibitem{mz01} L.V. Litvin, H.-P. Tranitz, W. Wegscheider, and C.
Strunk, Phys.\ Rev.\ B {\bf 75}, 033315 (2007).

\bibitem{mz02} P. Roulleau \textit{et al.}, Phys.\ Rev.\ B {\bf 76},
161309(R) (2007).

\bibitem{ant} V.J. Goldman and B. Su, Science {\bf 267}, 1010
(1995).

\bibitem{any} D.V. Averin and J.A. Nesteroff: Phys.\ Rev.\ Lett.
{\bf 99}, 096801 (2007).

\bibitem{mz3} C.L. Kane, Phys.\ Rev.\ Lett. {\bf 90}, 226802 (2003).

\bibitem{mz4} K.T. Law, D. E. Feldman, and Y. Gefen Phys.\ Rev.\ B
{\bf 74}, 045319 (2006).

\bibitem{usprl} V.V. Ponomarenko and D.V. Averin, Phys.\ Rev.\
Lett. {\bf 99}, 066803 (2007).

\bibitem{n1} L. Saminadayar \textit{et al.}, Phys. Rev.\ Lett. {\bf
79}, 2526 (1997).

\bibitem{n2} R. de-Picciotto \textit{et al.}, Nature {\bf 389}, 162
(1997).

\bibitem{comp1} D.V. Averin and V. J. Goldman, Solid State Commun.
{\bf121}, 25 (2001).

\bibitem{comp2} S. Das Sarma, M. Freedman, and C. Nayak, Phys.\ Rev.\
Lett. {\bf 94}, 166802 (2005).

\bibitem{comp3} A. Stern, Ann.\ Phys. {\bf 323}, 204 (2008).

\bibitem{wen} X.G. Wen, Adv.\ Phys. {\bf 44}, 405 (1995).

\bibitem{usprb} V.V. Ponomarenko and D.V. Averin, Phys.\ Rev.\ B
{\bf 79}, 045303 (2009).

\bibitem{usprb05} V.V. Ponomarenko and D.V. Averin, Phys.\ Rev.\ B
{\bf 71}, 241308(R) (2005).

\bibitem{fcs} L.S. Levitov and G.B. Lesovik, JETP Lett.
{\bf 58}, 230 (1993).

\bibitem{sw} H.~Saleur and U. Weiss, Phys.\ Rev.\ B {\bf 63},
201302(R) (2001).

\bibitem{ba} P. Fendley, A.W.W. Ludwig, and H. Saleur, Phys.\ Rev.\
Lett. {\bf 74}, 3005 (1995); Phys. Rev. B {\bf 52}, 8934
(1995).

\bibitem{za} S. Ghoshal and A.B. Zamolodchikov, Int.\ J.\ Mod.\
Phys. {\bf A9}, 3841 (1994).

\bibitem{GR} I.S. Gradshteyn and I.M.  Ryzhik, {\it Table of
Integrals, Series, and Products}, (Academic Press, 2007).

\bibitem{nazarov} D.A. Bagrets and Yu.V. Nazarov, Phys.\ Rev.\ B
{\bf 67}, 085316 (2003).

\bibitem{feldman2} D.E. Feldman  {\it et al.}, Phys.\ Rev.\ B {\bf
76}, 085333 (2007).

\bibitem{3d1} A.V. Timofeev \textit{et al.}, Phys. Rev.\ Lett.
{\bf 98}, 207001 (2007).

\bibitem{3d2} Q. Le Masne \textit{et al.}, Phys. Rev.\ Lett.
{\bf 102}, 067002 (2009).

\bibitem{3d3} G. Gershon \textit{et al.}, Phys. Rev.\ Lett.
{\bf 101}, 016803 (2008).

\bibitem{c3} L.S. Levitov and M. Reznikov, Phys.\ Rev.\ B
{\bf 70}, 115305 (2004).

\bibitem{expm2} S. Roddaro \textit{et al.}, Phys. Rev.\ Lett.
{\bf 103}, 016802 (2009).


\end{thebibliography}
\end{document}